\documentclass[twocolumn,letterpaper,amsmath,amssymb,floatfix,aps,superscriptaddress]{revtex4}

\usepackage{graphicx}
\usepackage{dcolumn}
\usepackage{bm}
\usepackage[usenames]{color}
\usepackage{hyperref}
\usepackage{ulem}

\begin{document}

\title{Dynamics of end-pulled polymer translocation through a nanopore}

\author{Jalal Sarabadani}
\email{jalal.sarabadani@aalto.fi}
\affiliation{Department of Applied Physics and COMP Center of Excellence, Aalto University School of Science, 
P.O. Box 11000, FI-00076 Aalto, Espoo, Finland}
\affiliation{School of Nano Science, Institute for Research in Fundamental Sciences (IPM), 19395-5531, Tehran, Iran}

\author{Bappa Ghosh}
\affiliation{Department of Chemistry, Indian Institute of Science Education and Research, Pune, Maharashtra, India}

\author{Srabanti Chaudhury}
\email{srabanti@iiserpune.ac.in}
\affiliation{Department of Chemistry, Indian Institute of Science Education and Research, Pune, Maharashtra, India}

\author{Tapio Ala-Nissila}
\affiliation{Department of Applied Physics and COMP Center of Excellence, Aalto University School of Science, 
P.O. Box 11000, FI-00076 Aalto, Espoo, Finland}
\affiliation{Departments of Mathematical Sciences and Physics, Loughborough University,
Loughborough, Leicestershire LE11 3TU, UK}

\begin{abstract}
We consider the translocation dynamics of a polymer chain forced through a nanopore by an external
force on its head monomer on the {\it trans} side. For a proper theoretical treatment we generalize 
the iso-flux tension propagation (IFTP) theory to include friction arising from the {\it trans} side 
subchain. The theory reveals a complicated scenario of multiple scaling regimes depending on the 
configurations of the {\it cis} and the {\it trans} side subchains. In the limit of high driving
forces $f$ such that the {\it trans} subchain is strongly stretched, the theory is in excellent 
agreement with molecular dynamics simulations and allows an exact analytic solution for the scaling 
of the translocation time ${\tau}$ as a function of the chain length $N_0$ and $f$. In this regime 
the asymptotic scaling exponents for ${\tau} \sim N_0^{\alpha} f^{\beta}$ are $\alpha=2$, and $\beta =-1$.
The theory reveals significant correction-to-scaling terms arising from the {\it cis} side subchain 
and pore friction, which lead to a very slow approach to $\alpha =2$ from below as a function
of increasing $N_0$.
\end{abstract}

\maketitle

{\it Introduction.} --
Dynamics of polymer translocation through nanopores has become an active research topic in soft matter and biological 
physics during the last two decades \cite{Tapio_review,Milchev_JPCM,Muthukumar_book} since the seminal experiment by Kasianowicz 
{\it et al.} in 1996 \cite{kasi1996}, and has 
many potential technological applications in biology, engineering and medicine just to name a few. 
In particular, translocation based setups have been suggested as 
rapid and inexpensive methods for DNA and other biopolymer sequencing. Motivated by these applications many experimental and 
theoretical \cite{meller2003,branton_PRL_2003,meller_biophys_j_2004,storm2005,branton2008,schadt2010,dimarzio1979,sung1996,%
muthu1999,chuang2001,metzler2003,kantor2004,grosberg2006,dubbeldam2007,sakaue2007,luo2008,sakaue2008,luo2009,bhatta2009,%
sakaue2010,rowghanian2011,saito2011,saito2012a,saito2012b,dubbeldam2011,ikonen2012a,ikonen2012b,ikonen2012c,ikonen2013,jalal2014,%
jalal2015,jalal2017,unbiased_Slater_1,unbiased_Slater_2,unbiased_Slater_3,unbiased_Polson,dubbeldam2013,Keyser_NatPhys_2006,%
Keyser_NatPhys_2009,Sischka,Bulushev_NanoLett_2014,Bulushev_NanoLett_2015,Bulushev_NanoLett_2016,Menais_SciRep_2016,Santtu_EPJE_2009,Huopaniemi_2007,Panja_2008}
studies have focused on the nature of the translocation dynamics. The three simplest basic translocation scenarios correspond to the cases of  
pore-driven, end-pulled, and unbiased setups \cite{,unbiased_Slater_1,unbiased_Slater_2,unbiased_Slater_3,unbiased_Polson}. 
In the pore-driven case an electric field, due to a voltage bias between the two sides 
of the membrane, acts on the monomers inside the pore. On the other hand, in the end-pulled case the polymer is pulled 
through a nanopore by either an optical or a magnetic tweezer \cite{Keyser_NatPhys_2006,Keyser_NatPhys_2009,Sischka,%
Bulushev_NanoLett_2014,Bulushev_NanoLett_2015,Bulushev_NanoLett_2016}. End-pulled translocation has been suggested to be
a good method to slow down and control 
the translocation process which is crucial for proper identification of the nucleotides in DNA 
sequencing \cite{kantor2004,Huopaniemi_2007,Santtu_EPJE_2009}. 

\begin{figure*}[t]\begin{center}
    \begin{minipage}[b]{0.28\textwidth}
    \begin{center}
        \includegraphics[width=1.0\textwidth]{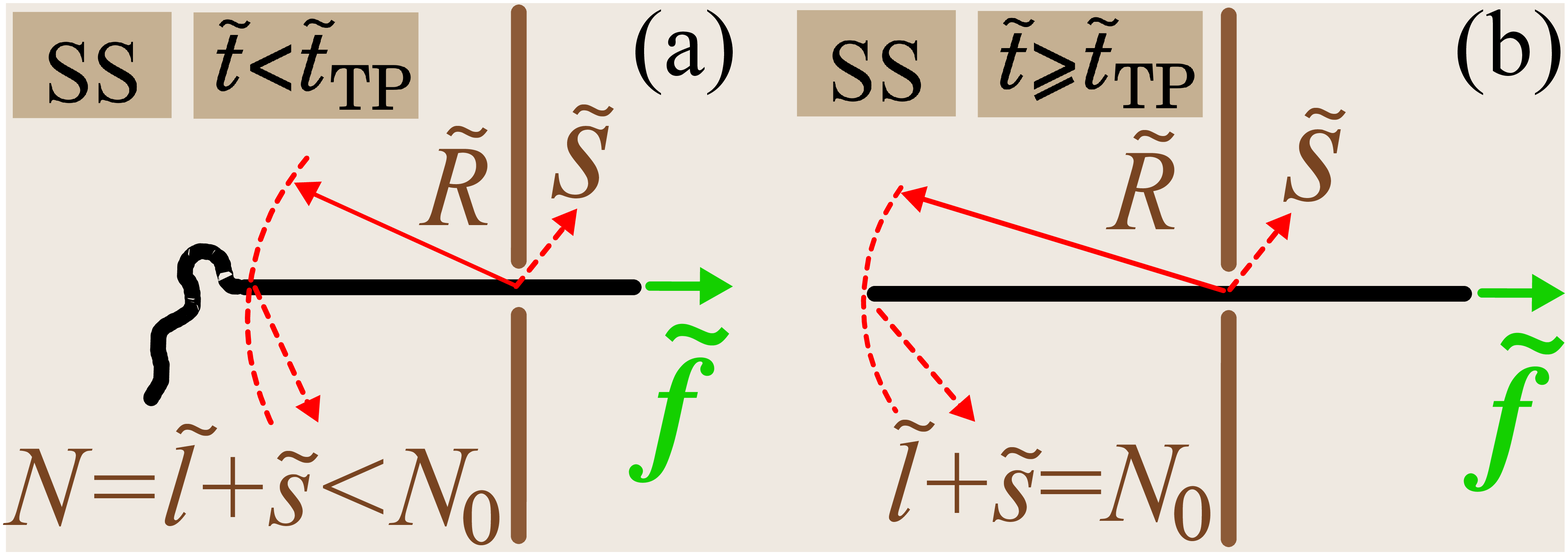}
    \end{center}\end{minipage} \hskip0.0cm
    \begin{minipage}[b]{0.28\textwidth}
    \begin{center}
        \includegraphics[width=1.0\textwidth]{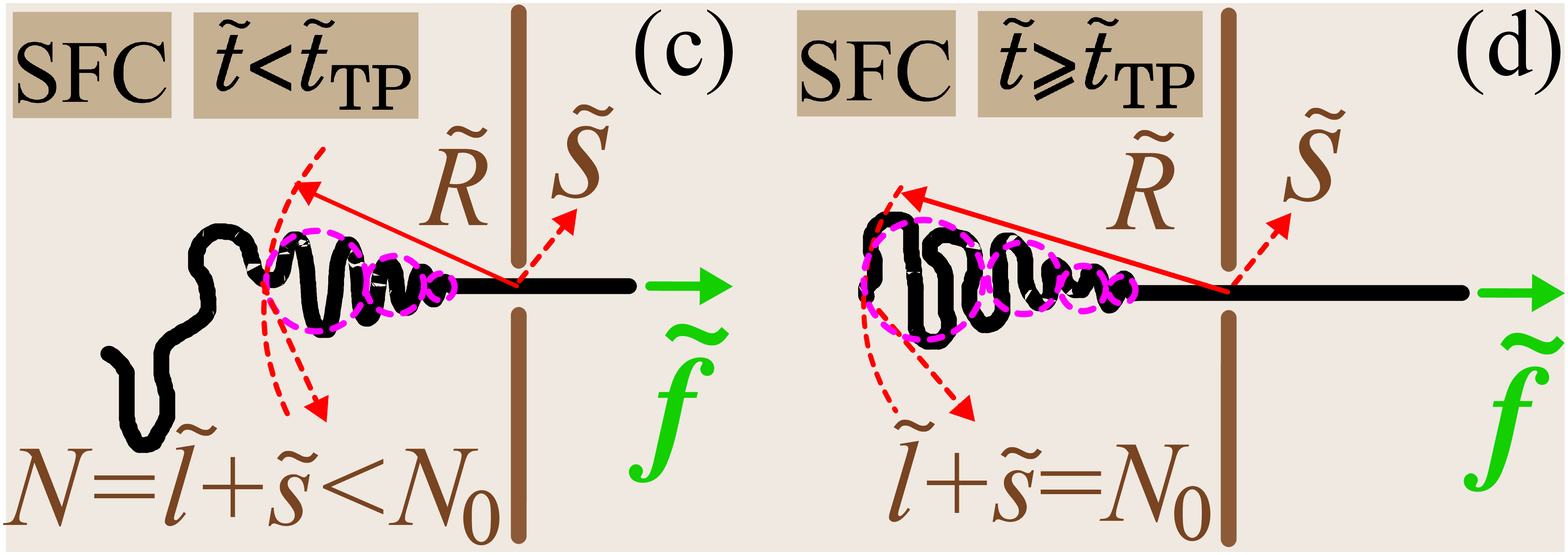}
    \end{center}\end{minipage} \hskip0.0cm
    \begin{minipage}[b]{0.28\textwidth}
    \begin{center}
        \includegraphics[width=1.0\textwidth]{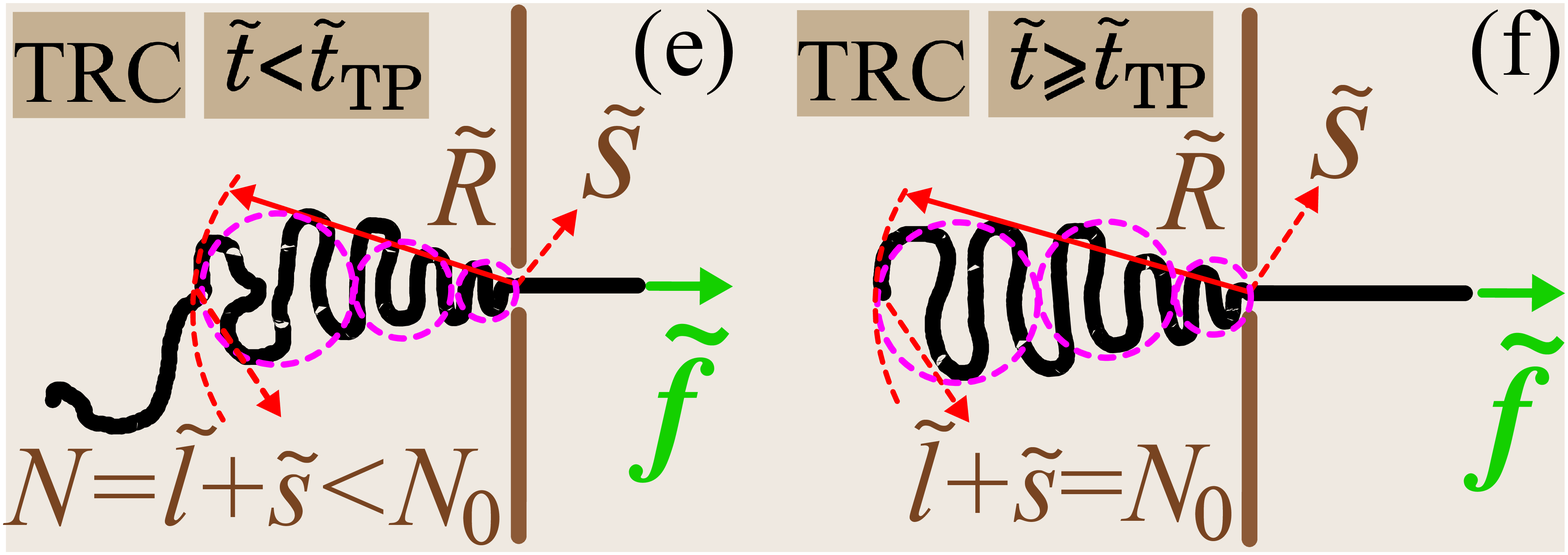}
    \end{center}\end{minipage}
\caption{Schematic of the various possible translocation scenarios during the tension propagation (TP) stage for the {\it trans} side 
strong stretching (SST) regime. The driving force $\tilde{f}$ acts only on the head monomer of the polymer in the {\it trans} side. 
The length of polymer is $N_0$ 
and the number of beads that have been already translocated into the {\it trans} side is denoted by $\tilde{s}$. The number of 
total beads influenced by the tension is $N= \tilde{l} + \tilde{s}$ (in the TP stage $N < N_0$). In panel (a) the {\it cis} 
side subchain is also in the SS regime (SSC) during the TP stage.
(b) The translocation process when the tension front reaches the end of the chain on the {\it cis} side in the SSC regime
(post propagation stage (PP) where $N = N_0$).
Panels (c) and (e) are the same as panel (a) but for the stem-flower (SFC) and trumpet (TRC) regimes in the {\it cis} side, respectively.
Panels (d) and (f) are the same as panel (b) but for SFC and TRC, respectively. See text for details.
} 
\label{fig:schimatic-SS-SFC-TRC}
\end{center}
\end{figure*}

Over the last few years a quantitative theory for pore-driven translocation dynamics has been developed \cite{jalal2014,jalal2015,jalal2017,
ikonen2012a,ikonen2012b,ikonen2012c,ikonen2013} based on the idea of tension
propagation by Sakaue \cite{sakaue2007}. The key role in this iso-flux \cite{rowghanian2011} tension propagation (IFTP) theory
is played by the time-dependent friction of the {\it cis} side subchain that resists the driving and leads to different regimes
of translocation dynamics depending on the strength of the driving force. Most recently, the theory has been extended
to the case of semi-flexible polymer chains, where there's an additional time-dependent frictional term arising from the {\it trans} side of the
chain due to chain stiffness \cite{jalal2017}. To date the case of end-pulled translocation has not been theoretically
treated beyond simple scaling arguments, however. Driven translocation processes are fundamentally far-from-equilibrium
phenomena and scaling arguments alone cannot capture the relevant physics. To this end, in this work we present a proper theoretical treatment
of the end-pulled setup based on the IFTP theory and augmented with  
extensive molecular dynamics (MD) simulations \cite{jalal2014,jalal2017,ikonen2012a,ikonen2012b,ikonen2012c}.
We show how the IFTP theory reveals the presence of translocation regimes that have not been previously
considered. In the present work we concentrate in the limit of strong driving, and derive an exact scaling 
formula for the scaling of the translocation time with the chain length and the driving force. 
The theory is shown to be in excellent agreement with MD simulations of coarse-grained polymer chains.


{\it Theory.} -- 
The IFTP theory has revealed that for the pore-driven case there are three different regimes of translocation dynamics corresponding to
different force strengths, namely the strong stretching (SS) limit of high forces, the stem-flower (SF) regime of intermediate forces, and
the trumpet (TR) regime of weak forces. For end-pulled chains this means that there is a complicated scenario of multiple regimes 
corresponding to different combinations of the chain configurations both on the {\it cis} and {\it trans} sides. For simplicity, in the present
work we will consider only the case of high driving forces $\tilde{f} \gtrsim N_0$, where $\tilde{f}$ is the pulling force 
and $N_0$ is the chain length, such that the {\it trans} side subchain is fully 
straightened at all times during the translocation process which we call the {\it trans}-SS scenario here. Correspondingly,
the {\it cis} side subchain can either be in the SS (SSC), stem-flower (SFC) or trumpet (TRC) regime.
The other regimes will be discussed in future work.

In Fig. \ref{fig:schimatic-SS-SFC-TRC} we show the
translocation regimes corresponding to the three different possible {\it cis} side subchain configurations.
In the {\it trans}-SS regime it is sufficient to use the deterministic limit of the iso-flux Brownian dynamics tension propagation theory 
without the entropic force term
\cite{ikonen2012a,ikonen2012b,jalal2014,jalal2015,jalal2017}.
The corresponding equation of motion for $\tilde{s}$, the translocation coordinate 
which is the number of beads in the {\it trans} side, is given by 
\footnote{Following our previous works dimensionless units denoted by tilde 
as $\tilde{X} \equiv X / X_u$ are used, with the units of time $t_u \equiv \eta \sigma^2 / (k_B T)$, 
length $s_u \equiv \sigma$, velocity $v_u \equiv \sigma/t_u = k_B T/(\eta \sigma)$, friction $\Gamma_u \equiv \eta$,
force $f_u \equiv k_B T/\sigma$, and monomer flux $\phi_u \equiv k_B T/(\eta \sigma^2)$,
where $T$ is the temperature of the system, $k_B$ is the Boltzmann constant, $\sigma$ is the segment length, and 
$\eta$ is the solvent friction per monomer.
The quantities without the tilde, such as the force, velocity, friction and length,
are expressed in Lennard-Jones units.}
\begin{equation}
\tilde{\Gamma} (\tilde{t}) \frac{d \tilde{s}}{ d \tilde{t}} =
\tilde{f},
\label{BD_equation_1}
\end{equation}
where $\tilde{\Gamma} (\tilde{t})$ is the effective friction, 
and $\tilde{f}$ is the external driving force. We have recently extended the IFTP theory to the case of pore-driven
semi-flexible chains \cite{jalal2017} and shown that the effective total friction $\tilde{\Gamma} (\tilde{t})$ must be written as 
$\tilde{\Gamma} (\tilde{t}) = \tilde{\eta}_{\rm cis} (\tilde{t}) + \tilde{\eta}_{\textrm{p}} + \tilde{\eta}_{\textrm{TS}} (\tilde{t})$,
where the first two terms $\tilde{\eta}_{\rm cis} (\tilde{t})$ and $\tilde{\eta}_{\textrm{p}} (\tilde{t})$ are the {\it cis} side subchain
and pore friction, respectively, and $\tilde{\eta}_{\textrm{TS}} (\tilde{t})$ is a new time dependent
{\it trans} side friction that cannot be absorbed in the constant pore friction. We expect the {\it trans} side friction to play
an important role for end-pulled polymers, too, and it needs to be explicitly taken into account.

\begin{figure*}[t]
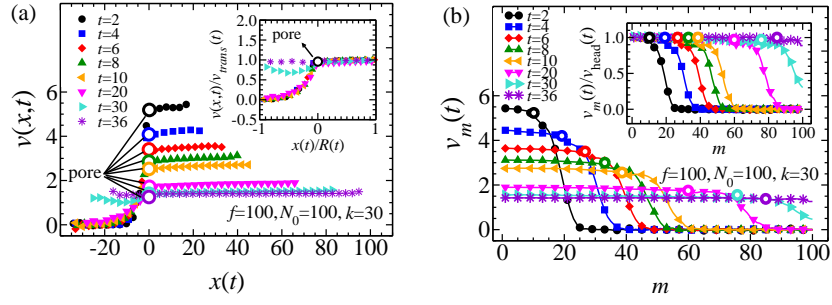
\begin{center}
    \begin{minipage}[b]{0.285\textwidth}\begin{center}
            \includegraphics[width=1.0\textwidth]{figure2a.eps}
    \end{center}\end{minipage} \hskip+0.5cm
    \begin{minipage}[b]{0.29\textwidth}\begin{center}
            \includegraphics[width=1.0\textwidth]{figure2b.eps}
    \end{center}\end{minipage}
\caption{(a) Velocity perpendicular to the wall in the {\it trans} side and towards the pore in the {\it cis} side for individual monomers $v (x,t)$ 
as a function of the distance $x (t) $ at different times ${t}=2- {36}$. The empty circles show the pore locations at different times.  
Inset shows normalized velocity of individual monomers $v (x, t) / v_{\rm trans} (t)$ 
as a function of the normalized distance $x (t) / R (t)$.
Here, $v_{{\rm trans}} (t)$ is the average velocity of the {\it trans} side sub-chain monomers and $R(t)$ is the amplitude of the tension front distance 
from the pore. The driving force acting on the head monomer is ${f} = 100$, the chain length is $N_0 = 100$, and the location of the pore is 
denoted by an open black circle.
(b) Velocity perpendicular to the wall in the {\it trans} side and towards the pore in the {\it cis} side of the monomers $v_{m} (t)$ 
as a function of the monomer number $m$ at different 
times ${t}=2-36$, where the empty colored circles show which monomer is inside the pore at each time. 
Inset shows the monomer velocity normalized by the head monomer velocity 
$v_{m} (t) / v_{\textrm{head}} (t)$ as a function of $m$. The values $m=1$ and 100 
denote the head and tail monomers, respectively.}
\label{fig:velocity}
\end{center}
\end{figure*}

Within the IFTP theory the dynamics of the chain on the {\it cis} side is solved with the corresponding TP equations.
To derive these we use the iso-flux approximation \cite{rowghanian2011}, where the monomer flux 
$\tilde{\phi}\equiv d\tilde{s}/d\tilde{t}$ on the mobile domain of the {\it cis} and {\it trans} sides 
and through the pore is constant in space but evolves in time. The tension front
is located at distance $\tilde{x}=-\tilde{R}(\tilde{t})$ from the pore. Inside the mobile domain, the external driving force is mediated by the 
chain backbone from the head bead (head of polymer on which the external driving force acts) all the way to the pore at $\tilde{x}=0$ and 
then to the last mobile monomer $N$ located at the tension front (see Fig. \ref{fig:schimatic-SS-SFC-TRC}). 
The magnitude of the tension force at distance $\tilde{x}$ can be calculated by considering the force balance relation
$d \tilde{f}( \tilde{x}' ) = - \tilde{\phi} (\tilde{t}) d \tilde{x}'$
for the differential element $d \tilde{x}'$ that is located between $\tilde{x}'$ and $\tilde{x}' + d\tilde{x}'$. 
By integrating the force balance relation over the distance from the head monomer to the pore on the {\it trans} side and then from the pore
to $\tilde{x}$ on the {\it cis} side, the tension force is obtained as $\tilde{f}(\tilde{x},\tilde{t}) = 
\tilde{f}_0 - \tilde{\phi} (\tilde{t}) \tilde{x}$, where $ \tilde{f}_0  
\equiv \tilde{f} - \tilde{\eta}_{\textrm{p}} \tilde{\phi}(\tilde{t}) - \tilde{\eta}_{\textrm{TS}} \tilde{\phi}(\tilde{t}) $ 
is the force at the pore entrance on the {\it cis} side.

It is important to note that the tension front is always located on the {\it cis} side of the chain since the {\it trans}
side is subject to a constant driving force. The same results have been verified for weaker and stronger forces. 
To illustrate this, in Fig. \ref{fig:velocity}(a) we plot the 
velocity perpendicular to the wall in the {\it trans} side and towards the pore in the {\it cis} side for individual monomers $v (x, t)$ 
as a function of the distance $x (t)$ at different times {${t}=2-36$} during the translocation process, 
with ${f}=100$, spring constant $k=30$ and chain length $N_0 = 100$. 
Inset shows normalized velocity 
$v (x, t) / v_{{\rm trans}} (t)$ as a function of the normalized distance $x (t) / R (t)$, 
where $v_{{\rm trans}} (t)$ is the average monomer velocity of the {\it trans} side sub-chain
and $R(t)$ is the amplitude of the tension front distance from the pore. 
Moreover, in Fig. \ref{fig:velocity}(b) the velocity perpendicular to the wall {in the {\it trans} side and towards the pore in the {\it cis} side} 
for each monomer $v_{{m}} (t)$ 
is plotted as a function of the monomer number ${{m}}$ ($1 \leq {m} \leq N_0$) with the same 
parameters as in Fig. \ref{fig:velocity}(a) at different times $t=2- {36}$. As can be seen, in both panels (a) and (b)
the velocity of the {\it trans} side monomers is almost constant and drops immediately close to the pore entrance on the {\it cis} side 
{in the TP stage.} 
This velocity drop occurs because of the reorientation of the mobile part of the chain on the {\it cis} side due to pulling.
{In the PP stage as time passes the velocity of the {\it cis} side sub-chain increases and finally becomes equal to that of the {\it trans} 
side sub-chain velocity. The results of the SFC and TRC regimes 
are shown to be in better quantitative agreement with the MD results than those from the SSC regime due to the occurence 
of this velocity change, as can bee seen in the waiting time distributions of Fig. 3.

In the SSC, SFC and TRC regimes the force balance equation is integrated over the mobile domain \cite{jalal2014} 
and the monomer flux is obtained as
\begin{equation}
\tilde{\phi} (\tilde{t}) = \frac{\tilde{f}}
{ \tilde{R} (\tilde{t}) +\tilde{\eta}_{\textrm{p}} + \tilde{\eta}_{_\textrm{TS}}^{\textrm{J}} },
\label{phi_equation}
\end{equation}
where the superscript J refers to the SS, SFC or TRC regimes from hereon.
By combining Eqs. \eqref{BD_equation_1} and \eqref{phi_equation}, 
the effective friction can be written as
\begin{equation}
\tilde{\Gamma} (\tilde{t}) = \tilde{R}(\tilde{t}) + \tilde{\eta}_{\textrm{p}} + \tilde{\eta}_{_\textrm{TS}}^{\textrm{J}},
\label{Gamma_equation}
\end{equation}
where in the high force limit $\tilde{\eta}_{_\textrm{TS}}^{\textrm{J}} = \tilde{s}$. This equation is formally
the same as that for the pore-driven semi-flexible case \cite{jalal2017}, but the {\it trans} side friction term
is different. In the present case it simply equals the distance of 
the head monomer of the chain from the pore because the {\it trans} side subchain is fully straightened here.

To determine the full solution for the time evolution of $\tilde{s}$ by Eqs. \eqref{BD_equation_1}, \eqref{phi_equation} 
and \eqref{Gamma_equation}, the position of the tension front $\tilde{R}(\tilde{t})$ must be known. 
To find this it should be noted that $\tilde{R} (\tilde{t})$ is 
equivalent to the root mean square of the end-to-end distance of the flexible chain $\tilde{R}_N = A_{\nu} N^{\nu}$,
where $A_{\nu} =1.15$ and $\nu=0.588$ is the 3D Flory exponent. Therefore, using $\tilde{R}_N$
together with mass conservation which implies $N = \tilde{l}_{\textrm{J}} + \tilde{s} <  N_0$ and 
$\tilde{l}_{\textrm{J}} + \tilde{s}= N_0$ in the TP and post propagation (PP) stages, respectively, we separately derive 
equations of motion for $\tilde{R}(\tilde{t})$ in the TP and PP stages.
Here, the number of mobile beads on the {\it cis} side is defined as 
$\tilde{l}_{\textrm{J}} = \int_{\tilde{x}=0}^{\tilde{R} (\tilde{t})} \tilde{\sigma}_{\textrm{J}} (\tilde{x} , \tilde{t}) d \tilde{x} $,
where the integration of the monomer number density $\tilde{\sigma}_{\textrm{J}} (\tilde{x} , \tilde{t} )$ 
is performed over the distance from the pore at $ \tilde{x} = 0$ to the tension front at $ \tilde{x} = \tilde{R} (\tilde{t})$.
The monomer number density is unity when the chain is straightened, and according to the blob theory it is given by
$\tilde{\sigma} (\tilde{x}) = |\tilde{f} (\tilde{x})|^{(\nu -1)/\nu}$ when the chain has the form of either a trumpet or a flower 
\cite{jalal2014,ikonen2012a}. The quantity
$\tilde{l}_{\textrm{J}}$ for different regimes can be obtained as \cite{jalal2014} $\tilde{l}_{\textrm{SS}} = \tilde{R} (\tilde{t})$, 
$\tilde{l}_{\textrm{SFC}} = \tilde{R} (\tilde{t}) + (1-\nu) \tilde{\phi}(\tilde{t})^{-1} /(2\nu -1) $, and 
$\tilde{l}_{\textrm{TRC}} = \nu \tilde{\phi}(\tilde{t})^{(\nu-1)/\nu} \tilde{R}(\tilde{t})^{(2\nu-1)/\nu } / (2\nu-1) $.
The time evolution of the tension front in the TP stage can be expressed as
\begin{equation}
\dot{\tilde{R}} (\tilde{t}) \!=\!
\frac{ \nu A_{\nu}^{1/\nu} \tilde{R}^{(\nu-1)/ \nu}  {\cal{H}}_{\textrm{J}} }
{ 1 + \nu A_{\nu}^{1/\nu} \tilde{R}^{(\nu-1)/ \nu} \tilde{\phi}(\tilde{t}) {\cal{L}}_{\textrm{J}} },
\label{R-dot_PT_SS_SFC_TRC}
\end{equation}
and in the PP stage as
\begin{equation}
\dot{\tilde{R}} (\tilde{t}) \!=\!
\frac{ {\cal{H}}_{\textrm{J}}  }
{ \tilde{\phi}(\tilde{t}) {\cal{L}}_{\textrm{J}}  },
\label{R-dot_PP_SS_SFC_TRC}
\end{equation}
\begin{figure}[b]\begin{center}
    \begin{minipage}[b]{0.28\textwidth}\begin{center}
        \includegraphics[width=1.0\textwidth]{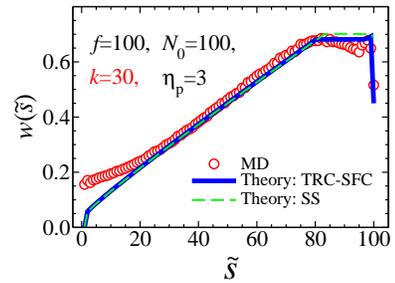}
    \end{center}\end{minipage}
\caption{The waiting time distribution $w (\tilde{s})$ as a function of the translocation coordinate $\tilde{s}$.
Here, the driving force is $f = 100$, spring constant in the MD simulations is $k=30$, $N_0 = 100$ is the chain length
and the pore friction in the theory is $\eta_{\textrm{p}} = 3$. The red circles show the MD simulation results while the 
solid blue and the dashed green lines are the results from the IFTP theory for the combination of the TRC and SFC regimes, and for the SSC regime, 
respectively.
} 
\label{fig:waiting-time_SS-SFC}
\end{center}
\end{figure}
where the operators
${\cal{L}}_{\textrm{SS}} = - \tilde{\phi} (\tilde{t})^{-1}$, 
${\cal{L}}_{\textrm{SFC}} = (\nu -1) [ \tilde{R} (\tilde{t}) + \tilde{\eta}_{\textrm{p}} +\tilde{s} ]^{-1} \tilde{\phi}(\tilde{t})^{-2}  /(2\nu-1) - \tilde{\phi} (\tilde{t})^{-1}  $,
${\cal{L}}_{\textrm{TRC}} = \tilde{\phi}(\tilde{t})^{-(1+\nu)/\nu} \tilde{R}(\tilde{t})^{(\nu-1)/\nu} 
[\tilde{R}(\tilde{t}) + \tilde{\eta}_{\textrm{p}} + \tilde{s} ]^{-1}
[ \tilde{\phi}(\tilde{t}) \tilde{R}(\tilde{t}) (\nu-1) /(2\nu-1) - \tilde{f} ]$,
${\cal{H}}_{\textrm{SS}} = \tilde{\phi}(\tilde{t})$,
${\cal{H}}_{\textrm{SFC}} = - \tilde{\phi}(\tilde{t})^2 {\cal{L}}_{\textrm{SFC}}$, and
${\cal{H}}_{\textrm{TRC}} = - \tilde{\phi}(\tilde{t})^2 {\cal{L}}_{\textrm{TRC}} 
+ \tilde{\phi}(\tilde{t}) - \tilde{\phi}(\tilde{t}) [\tilde{\phi}(\tilde{t}) \tilde{R}(\tilde{t})]^{(\nu-1)/\nu}$
corresponding to the three different regimes in Fig. 1.
To find the full solution of the IFTP theory, in the TP stage Eqs. (\ref{BD_equation_1}), (\ref{phi_equation}), (\ref{Gamma_equation}) and 
(\ref{R-dot_PT_SS_SFC_TRC}), and in the PP stage Eqs. (\ref{BD_equation_1}), (\ref{phi_equation}), 
(\ref{Gamma_equation}) and (\ref{R-dot_PP_SS_SFC_TRC}) 
must be self-consistently solved.

{\it Waiting time distribution.} --
To test the validity of the IFTP theory for the end-pulled case in Fig. \ref{fig:waiting-time_SS-SFC} the monomer waiting time 
distribution $w (\tilde{s})$ (the time that each bead spends at the pore) is plotted as a 
function of the translocation coordinate $\tilde{s}$. The red circles show our MD data obtained using the same bead-spring model
as in our previous works (the details can be found in the Supplementary). The solid blue line comes from the IFTP 
theory when the equations of motion are solved with a combination of the SFC and TRC regimes. This is determined based on
the value of the force at the pore entrance in the {\it cis} side, namely if $\tilde{f}_0 \gtrsim 1$ 
the equations must be solved in the SFC regime while for $\tilde{f}_0 \lesssim 1$ we solve the equations in the TRC regime.
The dashed green line represents the 
waiting time when the IFTP theory is solved in the SSC regime. As can be seen in Fig. \ref{fig:waiting-time_SS-SFC}, the combination of 
SFC and TRC matches perfectly with the MD data even in the PP stage. 
The dashed green curve for the SSC regime overestimates the waiting 
time in the PP stage because of the reorientation of the mobile part by the pore on the {\it cis} side in the MD simulations. 
Moreover, for small values of $\tilde{s}$
the solution of the IFTP theory underestimates the waiting time. This occurs in part because of bond stretching and also due to 
the beginning of the reorientation of the mobile part on the {\it cis} side as discussed in the theory subsection.
As this discrepancy exists only for small values of $\tilde{s}$ and the main contribution of the waiting time 
and consequently the translocation time comes from the larger $\tilde{s}$ almost at the end of the TP stage, and over the whole PP stage, 
the IFTP theory correctly predicts the overall behavior of the translocation process.

{\it Scaling exponents for translocation.} --
A fundamental characteristic of translocation dynamics is the scaling of the translocation time $\tilde{\tau}$ 
as a function of the chain length and the external driving force 
$\tilde{\tau} \propto \tilde{f}^{\beta} N_0^{\alpha}$, where $\alpha$ is the translocation exponent and $\beta$ is the force exponent.
\begin{figure}[b]\begin{center}
    \begin{minipage}[b]{0.28\textwidth}\begin{center}
        \includegraphics[width=1.0\textwidth]{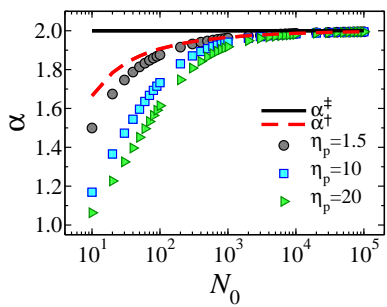}
    \end{center}\end{minipage}
\caption{The effective translocation time exponent $\alpha$ as a function of chain length $N_0$ for various values of the pore friction 
$\eta_{\textrm{p}}=1.5$ (black circles), 10 (blue squares) and 20 (green triangles). The dashed red and the horizontal 
black solid lines represent the rescaled translocation exponents $\alpha^{\dag}$ and $\alpha^{\ddag}$, respectively.}
\label{fig:translocation-exponent_SS-SFC-TR}
\end{center}
\end{figure}
Following Refs. \cite{jalal2014,jalal2017} we can derive an exact analytic expression for $\tilde{\tau}$,
which is a sum of the TP ($\tilde{\tau}_\mathrm{tp}$) and PP ($\tilde{\tau}_\mathrm{pp}$) contributions 
to the translocation time as $\tilde{\tau} = \big[ \int_0^{N_0} \tilde{R}_N dN 
+ \tilde{\eta}_{\textrm{p}} N_0 \big]/\tilde{f} + \tilde{\tau}_{\textrm{TS}}^{\textrm{J}}$. Since
the {\it trans} side of the subchain is fully straightened for all the regimes of SSC, SFC and TRC (cf. Fig. 1),
$\tilde{\tau}_{\textrm{TS}}^{\textrm{J}} = N_0^2/(2\tilde{f})$ and therefore
\begin{equation}
\tilde{\tau} = \frac{1}{\tilde{f}} \bigg[ \frac{A_{\nu}}{\nu + 1} N_0^{\nu +1} + \tilde{\eta}_{\textrm{p}} N_0 + \frac{1}{2} N_0^2 \bigg].
\label{scaling_trans_time}
\end{equation}
Here, for the TP stage the conservation of mass is $N= \tilde{s}+ \tilde{l}$ and
the TP time can be obtained by integration of $N$ from $0$ to $N_0$, while in the 
PP stage the conservation of the mass gives $N= \tilde{s}+ \tilde{l}=N_0$ and
the PP time is solved by integration of $\tilde{R}$ from $\tilde{R}_{N_0}$ to zero.
As can be seen in Eq. \eqref{scaling_trans_time} the force exponent is $\beta = -1$.
The first two terms in Eq. \eqref{scaling_trans_time} are identical to the case of
a pore-driven flexible chain \cite{jalal2014}, and the new term proportional to
$N_0^2$ arises from the explicit {\it trans} side friction for the end-pulled case.
While the asymptotic scaling exponent is $\alpha=2$, the two
correction-to-asymptotic-scaling terms lead to pronounced crossover behavior due
to the {\it cis} side and pore friction. 
To quantify the crossover scaling behavior in the high force limit, in Fig. \ref{fig:translocation-exponent_SS-SFC-TR} we plot the 
translocation time exponents $\alpha$, $\alpha^{\dag}$, and $\alpha^{\ddag}$ as a function of $N_0$ for various
values of the bare pore friction. The
two latter exponents have been defined by subtracting the correction terms as
$\tilde{\tau}^{\dag}= \tilde{\tau} - \tilde{\eta}_{\textrm{p}} N_0 / \tilde{f} \sim N_0^{\alpha^\dag}$ and 
$\tilde{\tau}^{\ddag}= \tilde{\tau}   - \big[ \int_{0}^{N_0} \tilde{R}_N dN + \tilde{\eta}_{\textrm{p}} N_0 \big]/ \tilde{f}  \sim N_0^{\alpha^{\ddag}} $.
For these typical parameters, asymptotic scaling is seen for very long chains only.

{\it Summary.} --
We have studied the dynamics end-pulled polymer translocation through a nanopore by means of the proper analytical IFTP theory which reveals the
existence of a complicated scenario depending on the strength of the driving force. Even in the case where the force is high enough such
that the {\it trans} side of the chain is straight which is the focus here, 
there are three different regimes depending on the conformation of the {\it cis} side subchain.
We have derived the corresponding equations of motion for the tension front propagation explicitly including the {\it trans} side friction. The theory
shows excellent agreement with MD simulations for the waiting time distribution of the monomers, and allows an exact solution for the scaling
of the translocation time as a function of the chain length and the driving force. As expected from simple scaling arguments, the theory recovers
the asymptotic exponent $\alpha = 2$, but with significant correction-to-scaling terms that come from the {\it cis} subchain and pore friction. Asymptotic scaling
is recovered only for excessively long chains for typical parameter values.

{\it Acknowledgements.} -- This work has been supported in part by the Academy of Finland through its Centers of Excellence Program under project 
nos. 251748 and 284621. The numerical calculations were performed using computer resources from the Aalto University School of 
Science ``Science-IT'' project, and from CSC - Center for Scientific Computing Ltd. The MD simulations were performed with the LAMMPS and GROMACS
simulation softwares. S.C. and B.G. would like to thank IISER Pune and Centre for Development of Advanced Computing, Pune, India for providing the 
HPC facility. B.G. would like to acknowledge IISER Pune for fellowship. S. C. would like to thank DAE-BRNS, Grant No: (37(2)/14/08/2016-BRNS) for 
high  performance  computer  cluster.

\end{document}